\documentclass[11pt]{article}
\usepackage{moriond,epsfig}

\newcommand{\pt}        {\mbox{p$_{\mathrm T}$}}

\newcommand{\ttbar}     {\mbox{$t\bar{t}$}}
\newcommand{\ppbar}     {\mbox{$p\bar{p}$}}
\newcommand{\invfb}     {\mbox{fb$^{-1}$}}

\def\be{\begin{equation}}
\def\ee{\end{equation}}
\def\bea{\begin{eqnarray}}
\def\eea{\end{eqnarray}}

\begin{document}
\vspace*{4cm}
\title{Single top quark production at the Tevatron}

\author{ Ar\'an Garc\'\i a-Bellido }

\address{Department of Physics and Astronomy, University of Rochester, 500 Wilson Blvd., Rochester, NY 14627}

\maketitle\abstracts{The observation of single top quark production by the CDF and D0 collaborations is one of the flagship measurements of the Run II of the Tevatron. The Tevatron combined single top quark cross section is measured to be: $\sigma(tb+X,~tqb+X)=2.8^{+0.6}_{-0.5}$~pb for a top quark mass of 170~GeV. This result is in agreement with the standard model production of a single top quark together with a jet in $\ppbar$ collisions at $\sqrt{s}$=1.96~TeV and allows to measure the CKM matrix element $|V_{tb}|$ without assumptions about the number of quark families. Other analyses involving tau leptons have been performed, and several properties, like the top quark width or the polarization have been measured.}

\section{Introduction}
The production of top quarks at the Tevatron occurs mainly in $\ttbar$ pairs through the strong interaction with a cross section of $7.91\pm0.91$~pb~\cite{ttbarxs}, but top quarks can also be produced singly via the electroweak interaction with a cross section of $3.46\pm0.14$~pb~\cite{stxs} for $m_t=170$~GeV. Two production modes are dominant at the Tevatron, categorized by how the $W$ boson is exchanged: the $s$-channel $\ppbar\to tb+X$ and the $t$-channel $\ppbar\to tqb+X$.\cite{singletop-theory} This note describes the two analysis by CDF and D0 that observe for the first time the combined $s+t$ single top quark production, a D0 analysis that measures separately the $s$- and $t$-channels cross section, a CDF measurement of the single top cross section in the missing transverse energy (MET) and jets final state and another one from D0 that reconstructs the tau+jets final state, the first determination of the top quark width by D0 using the singletop, and finally the first measurement of the top quark polarization in single top events by CDF.  

\section{Single top quark observation}
The final state consists of one high \pt~lepton (electron or muon), missing energy, and at least two jets, one or two of them originating from $b$-quarks. Loose selections are employed by both CDF and D0 to select events with this final state. The main background is $W$+jets, specially $Wb\bar{b}$, $Wc\bar{c}$ and $Wcj$, which are normalized to data before $b$-tagging. Good agreement is achieved after $b$-tagging between the data and the predicted backgrounds. D0 selects 4,519 events in data, with 4,652$\pm$352 expected events (including 223$\pm$30 $s$+$t$ expected signal events) in 2.3~\invfb~of data. CDF selects 3,315 events in data, with 3,377$\pm$505 expected events (including 191$\pm$28 $s$+$t$ expected signal events) in 3.2~\invfb~of data.
Since there is no single variable that differentiates the signal from the large backgrounds, several multivariate techniques are employed and then combined by each experiment, separately for each channel: electron or muon, two, three (or four) jets, and 1 or 2 $b$-tagged jets. D0 uses three multivariate techniques: Boosted Decision Trees (BDT), Bayesian Neural Networks (BNN) and a Matrix Elements (ME) probability based calculation. Similarly, CDF uses BDTs, Neural Networks, ME, and likelihood functions. These different discriminants are run over the same data sample and since they are not 100\% correlated, but rather 60-90\%, the combination improves the sensitivity of any single discriminant. 

\begin{figure}[h!tb]
\begin{center}
\includegraphics[width=0.3\textwidth]{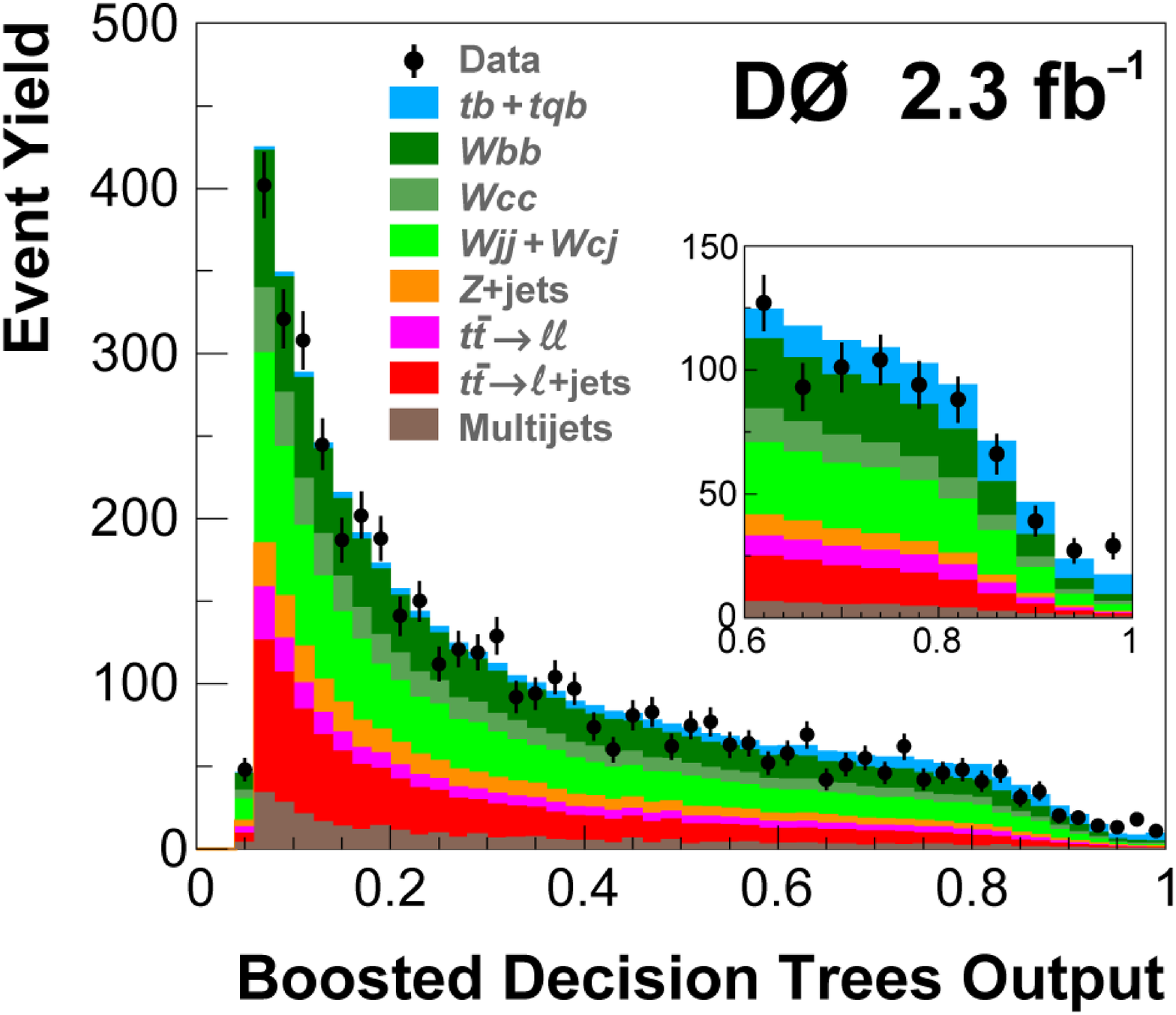}
\includegraphics[width=0.34\textwidth]{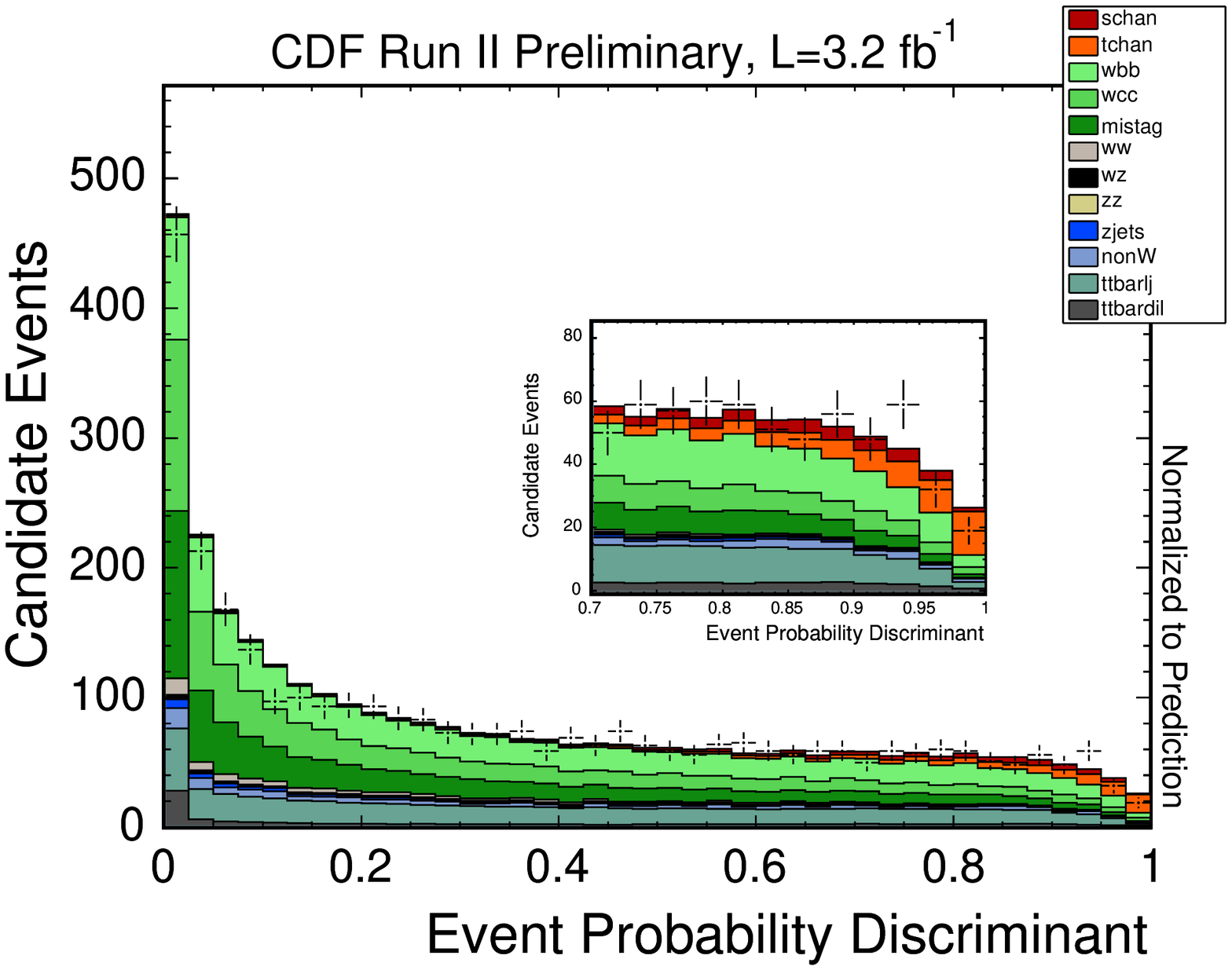}
\includegraphics[width=0.33\textwidth]{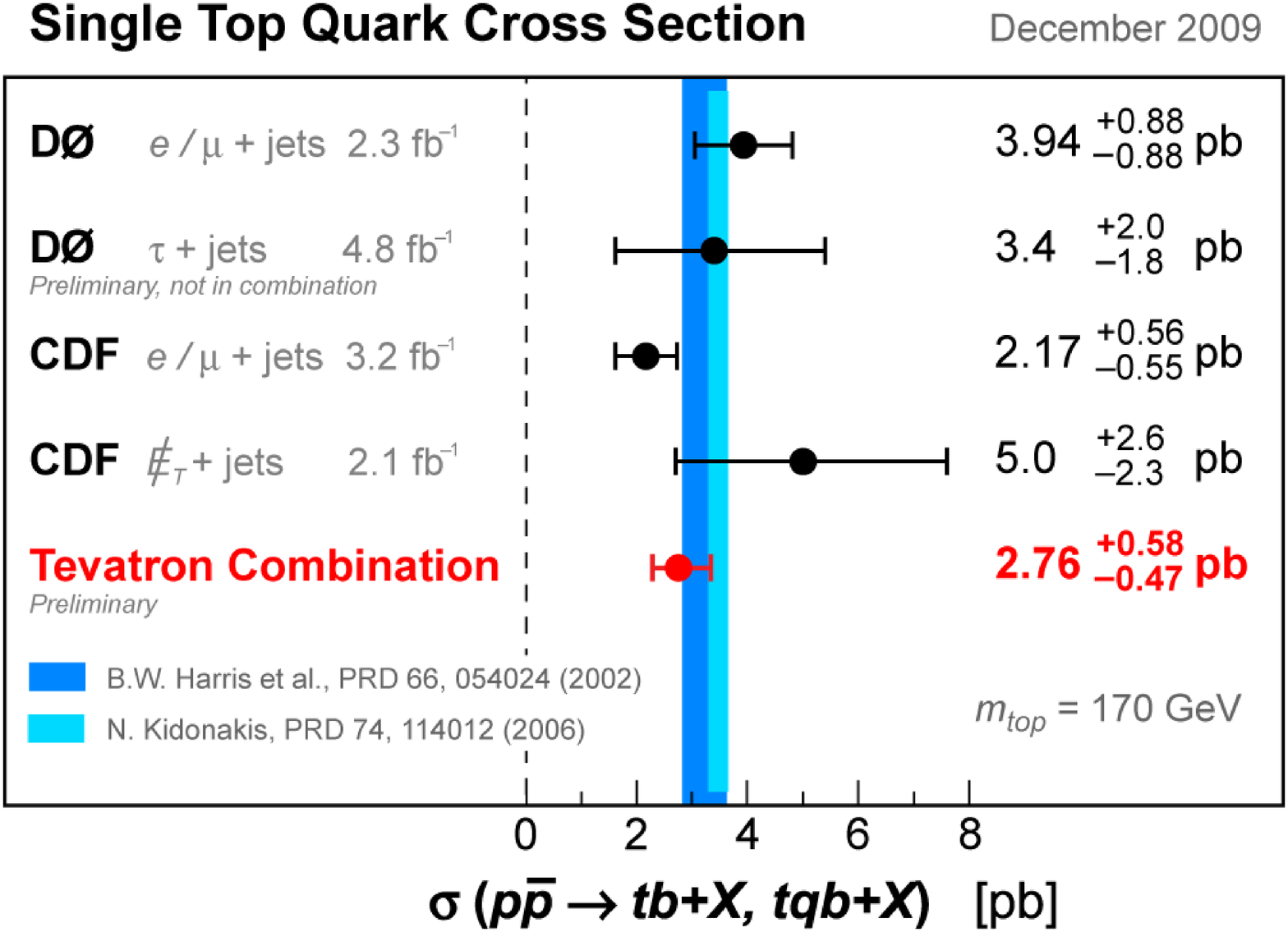}
\end{center}
\vspace{-0.1in}
\caption{Left: D0 BDT output. Middle: CDF ME output. Right: individual CDF and D0 results and Tevatron combination of the $s+t$ single top quark production cross section.}
\label{fig:combo}
\end{figure}

By combining their individual multivariate methods, both CDF and D0 have established separately the presence of single top quark production in their data with 5$\sigma$ in 3.2~\invfb~and 2.3~\invfb~respectively.\cite{observation} Additionally, the Tevatron collaborations have combined their analyses following the standard procedures of the Tevatron Electroweak Working Group.\cite{cdfd0comb} The result is a measured $s+t$ cross section of $2.76^{+0.58}_{-0.47}$~pb, which is converted in a measurement of the CKM matrix element: $|V_{tb}|=0.88 \pm 0.07$, equivalent to 8\% relative uncertainty on $|V_{tb}|$. Figure~\ref{fig:combo} shows the summary of the individual CDF and D0 results and the combined measurement.

\section{Single top quarks in $W\to\tau\nu$ decays}
D0 has searched in the $\tau$ final state with 4.8~\invfb~of integrated luminosity, directly reconstructing the hadronic $\tau$ candidates.\cite{d0taujets} The selection requires one hadronic tau candidate, missing transverse energy, and two or three jets with one or two $b$-tags. 
This analysis developed a new BDT technique to identify hadronic taus, it improved efficiency by $\approx5\%$ with respect the Neural Network usually employed by D0, for the same rejection rate of 98\%. In this final state, the main background is multijet production, which is derived from data. A final BDT is trained to discriminate against multijet background and its output used as discriminant to measure the single top cross section. The BDT output is shown in Figure~\ref{fig:s-t-sep}. The result from this analysis is a cross section of $3.4\pm2.0$~pb, or an upper limit of 7.3~pb at 95\%~C.L. This result was not part of the Tevatron combination described above. When combined with the D0 electron and muon final states, it yields the most precise measurement of the cross section of $3.84^{+0.89}_{-0.83}$~pb. 

CDF has published a similar search~\cite{cdfmetjets}, also looking at the tau final state, but not reconstructing the hadronic tau candidates directly nor electrons nor muons. The sample used consists of 2.1~\invfb~with MET and jets, so it is enriched in $W\to\tau\nu$ decays. Using two Neural Networks, one trained against the dominant QCD background, and another against the remaining backgrounds, the measured cross section is $4.9^{+2.5}_{-2.2}$~pb. This result was included in the CDF observation paper and is part of the Tevatron combination.

\section{Separate $s$- and $t$-channel search}
Using the same dataset of 2.3~\invfb~as for the observation, D0 has published a separate search for $s$ and $t$-channels.\cite{d0tchan} The motivation for this search lies in the fact that the observation analyses assumed the SM ratio for the relative contributions of $s$ and $t$ channels in the $s+t$ signal. That assumption is relaxed here, to probe possible new physics in each different channel. In general, new physics affect the $s$ and $t$-channel differently: new heavy bosons would enhance the $s$-channel production cross section, while anomalous couplings like flavor changing neutral currents, and CP violating or tensor couplings would affect the $t$-channel production cross section.\cite{Tait:2000sh} 
D0 employs the same multivariate techniques as in the observation, but now trained with $s$-channel as signal and $t$-channel as background, and viceversa. The outputs from the BDT, BNN and ME outputs are then combined with a Neural Network. The results are $3.14^{+0.94}_{-0.80}$~pb for the $t$-channel and 
$1.05 \pm 0.81$~pb for the $s$-channel. The measured $t$-channel result is found to have a significance of 4.8 standard deviations and is consistent with the standard model prediction. Figure~\ref{fig:s-t-sep} shows the two-dimensional plane of the $s$- and $t$-channels cross section, with possible new physics signatures. This result is still not sensitive enough to exclude new physics models, but with more data and improved selections it will be important to tell apart possible sources of new physics. 

\section{Top width from single top quarks}
Direct measurements of the top quark width from the invariant mass distribution are limited by the experimental resolution and statistics. D0 has performed an indirect measurement~\cite{d0width}, utilizing the $t$-channel single top cross section measurement~\cite{d0tchan} to extract the partial width $\Gamma(t\to Wb)$, and using the measurement of the ratio of branching fractions $R=\mathcal{B}(t\to Wb)/\mathcal{B}(t\to Wq)$ in $\ttbar$ decays to extract the branching fraction $\mathcal{B}(t\to Wb)$.\cite{rd0} The total width is then derived from the two separate measurements: $\Gamma_t=\Gamma(t\to Wb)/\mathcal{B}(t\to Wb)$. This method yields the most precise determination of the total width: $\Gamma_t=2.1\pm0.6$~GeV, or a lifetime of $\tau_t=(3 \pm 1)10^{-25}$~s, for $m_t = 170$~GeV. This result can be used to exclude some models of non-SM helicity amplitudes of the top quark, and constrain the coupling of a fourth generation heavy $b'$ quark with the $W$~boson and the top quark.

\section{Top quark polarization}
\begin{figure}[h!tb]
\begin{center}
\includegraphics[width=0.36\textwidth]{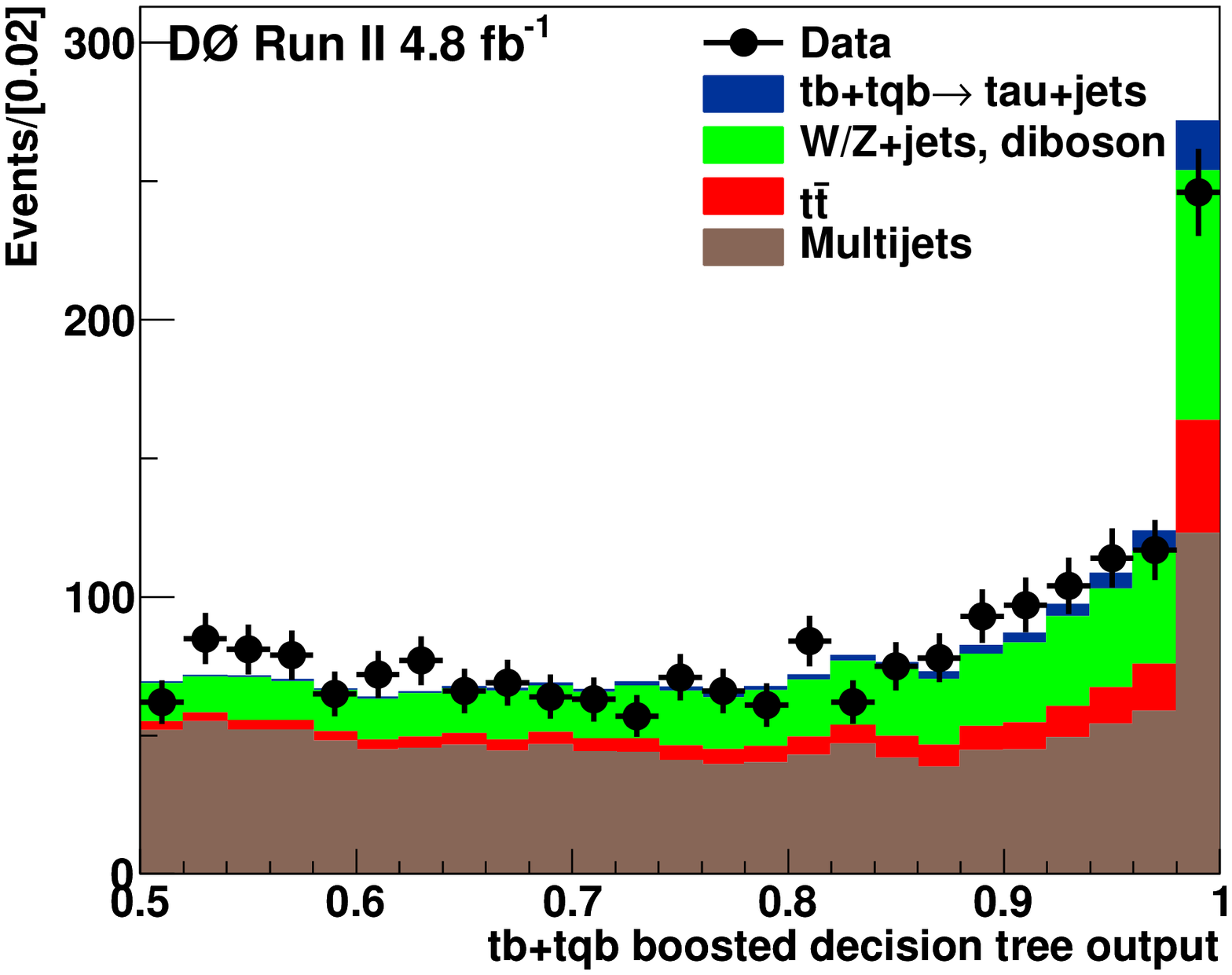}
\includegraphics[width=0.3\textwidth]{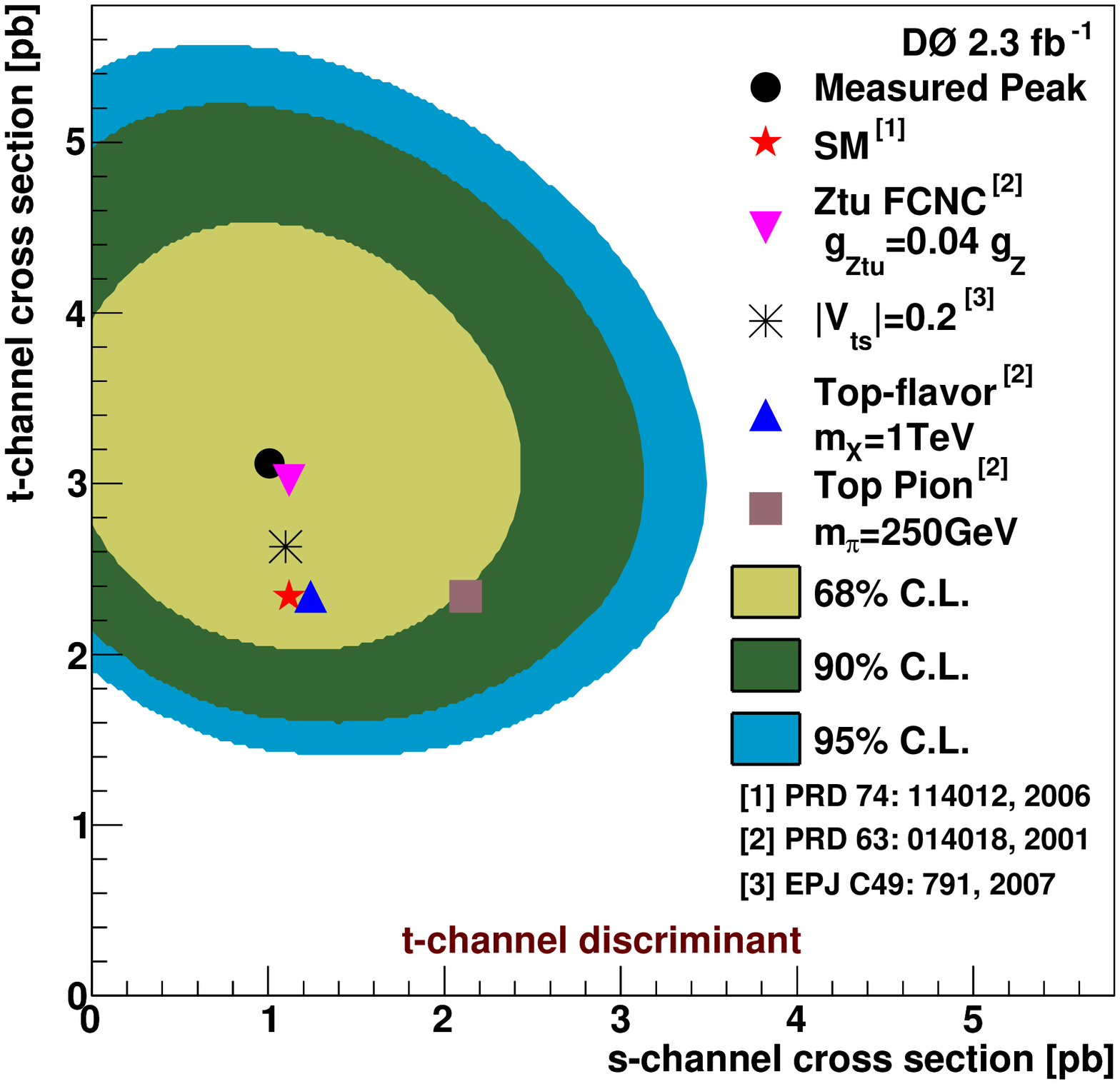}
\includegraphics[width=0.28\textwidth]{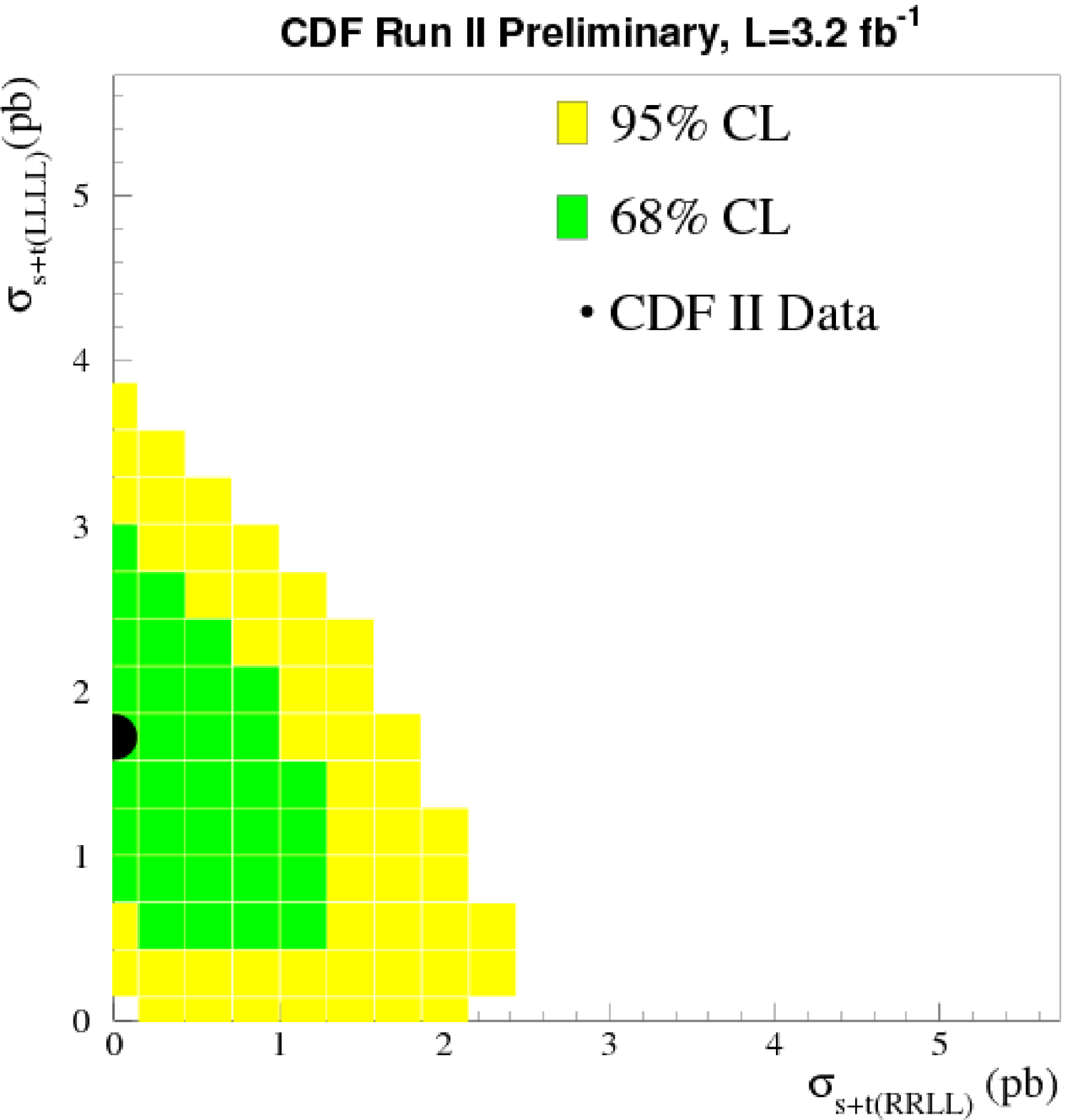}
\end{center}
\vspace{-0.1in}
\caption{Left: BDT output in the D0 $\tau$ channel measurement. Middle: $s$ vs. $t$-channel plane, with the separate D0 measurement and contributions from new physics. Right: CDF's measurement of the SM and the RRLL cross sections for the polarization.}
\label{fig:s-t-sep} 
\end{figure}

Single top quark production offers a nice sample of 100\% polarized top quarks along the accompanying down-type quark axis.
Non-SM contributions, such as $W'$~bosons, charged Higgs bosons, or flavor changing neutral currents, can change the polarization of top quarks. CDF has performed the first analysis~\cite{cdfpol} to measure the polarization assuming right-handed couplings in the production vertices and SM left-handed couplings in the decay, a model denoted by RRLL. The most discriminant variable is the angle between the lepton and the $b$-quark in the top rest frame. This variable has been added to the existing likelihood function from the observation: two discriminants were trained separately one for the SM production and another for the assumed RRLL production and decay. As seen in Fig.~\ref{fig:s-t-sep}, the best fit in the two dimensional plane of the measured cross sections is in agreement with the SM:  $\sigma_{s+t}(\mathrm{SM})=1.72$~pb and $\sigma_{s+t}(\mathrm{RRLL})=0$~pb. This can be translated into a measurement of the polarization: $\mathcal{P}=\frac{\sigma_{s+t}(\mathrm{RRLL})-\sigma_{s+t}(\mathrm{SM})}{\sigma_{s+t}(\mathrm{RRLL})+\sigma_{s+t}(\mathrm{SM})}=-1.0^{+1.5}_{-0}$. 

\section{Conclusions}
The observation in 2009 of single top quark production is a very significant milestone of the Tevatron program. This flagship measurement of Run II has opened the door to further studies of the properties of top quarks and searches beyond the standard model. We have presented here only a fraction of them, namely the measurements in hadronic $\tau$ final states, the extraction of the top quark width, and a measurement of the top quark polarization. More analyses can be found in the top group webpages from CDF and D0.\cite{webpages} 

\section*{Acknowledgments}
The author thanks the organizers of the Rencontres de Moriond for the
enjoyable and fruitful atmosphere of the meeting. His accommodation for this conference was funded by an NSF grant through Brown University.

\end{document}